# Splitting of levels in a circular dielectric waveguide

Nikolai I. Petrov

*Lenina Street, 19-39, Istra, Moscow region 143500, Russia*
*Corresponding author: petrovni@mail.ru*

A splitting of modes in a circular graded-index optical fiber is demonstrated by solving the full Maxwell equations using the perturbation analysis. It is shown that the degeneracy of vortex Laguerre-Gauss modes with distinct orbital angular momentum (OAM) and polarization (spin) but the same total angular momentum is lifted due to the spin-orbit (vector) and tensor forces. Numerical estimations of group delays of modes in optical fiber and frequency splitting in Fabry-Perot and ring resonators are presented.

Recent advances in fabrication technology have given rise to wavelength-scale optical systems, such as microcavities, microlasers, enhanced sensors, light-matter modulators and nanofibers [1-4]. Spin-orbit interaction in these systems may significantly modify the energy spectrum causing the splitting and degeneracy lifting of modes [5,6]. These effects influence on the group delays of modes in optical waveguides [7,8] and become important in fiber optic communications with high carrying capacities and faster transmission rates.

It is known that the problem of defining laser resonator modes is mathematically equivalent to the problem of finding propagation modes of a lens-like medium, where the Laguerre-Gauss (LG) functions are the modal solutions. Indeed, LG modes were predominately observed also in micro-cavities [3,4]. It was shown experimentally that mode profiles are radially symmetric LG modes, but they do not display the frequency degeneracies typical of large scale optical cavities [3,4]. Although the paraxial theory is able to model the cavity modes, and a qualitative agreement between theory and experiment was shown, the appropriate mode solutions can only be determined by solving the full Maxwell equations [5,9,10]. Modes of cylindrical waveguide in a scalar approximation are degenerated. When vector term in wave equation determining spin-orbit interaction is considered, the spectrum of propagation constant is split. The splitting of levels due to $\nabla\varepsilon$ term (spin-orbit interaction of photons) in lens-like media was considered in several papers [6-8].

In this paper the spectrum of modes of cylindrical optical fiber is determined analytically by solving the three-component field Maxwell's equations using the perturbation analysis. Polarization and nonparaxiality effects are considered in conjunction. The fine splitting of levels with the same total angular momentum and existence of hybrid modes with polarization-OAM entanglement are demonstrated. High sensitivity of the splitting to the fundamental mode size is shown.

The Maxwell equations for the electric field $\vec{E}\exp(-i\nu t)$ in a general inhomogeneous medium with dielectric constant $\varepsilon(x, y)$ reduce to:

$$\begin{aligned}(\nabla_\perp^2 + k^2 n^2)\vec{e}_\perp - i\beta\nabla_\perp e_z - \nabla_\perp \nabla_\perp \vec{e}_\perp &= \beta^2 \vec{e}_\perp \\ (\nabla_\perp^2 + k^2 n^2)e_z + i\beta\vec{e}_\perp \nabla_\perp \ln n^2 &= \beta^2 e_z\end{aligned} \quad (1)$$

where $k = 2\pi/\lambda$ is the wavenumber and $n^2(x,y)$ is the dielectric permittivity of the medium, $\beta$ is the propagation constant.

It is assumed that the dependence on time and $z$ is $\exp[-i(\nu t - \beta z)]$. Rewrite the system (1) in the form:

$$\left(\nabla_\perp^2 + k^2 n^2 + W_0 + \frac{1}{4}P^2\right)\vec{E} = \left(\beta - \frac{1}{2}P\right)^2 \vec{E}, \quad (2)$$

where $\vec{E} = \begin{pmatrix} e_x \\ e_y \\ e_z \end{pmatrix}$, $W_0 = \begin{pmatrix} -\frac{\partial^2}{\partial x^2} & -\frac{\partial^2}{\partial x \partial y} & 0 \\ -\frac{\partial^2}{\partial y \partial x} & -\frac{\partial^2}{\partial y^2} & 0 \\ 0 & 0 & 0 \end{pmatrix}$

$$P = \begin{pmatrix} 0 & 0 & -i\frac{\partial}{\partial x} \\ 0 & 0 & -i\frac{\partial}{\partial y} \\ \frac{i}{n^2}\frac{\partial n^2}{\partial x} & \frac{i}{n^2}\frac{\partial n^2}{\partial y} & 0 \end{pmatrix}.$$

By formally taking the square root of both sides, we have an equation which is equivalent to the stationary Schrodinger equation for the reduced field $\Psi(x, y)$:

$$\hat{H}\Psi = \varepsilon\Psi, \quad (3)$$

where $\varepsilon = \delta\beta = \beta - kn_0$ and $\Psi(x, y)$ are the eigenvalue and eigenfunction of the Hamiltonian, accordingly, $n_0 = n(0,0)$,

$$\hat{H} = \hat{H}^{(0)} + \hat{H}^{(1)} + \ldots, \qquad (4)$$

$$\hat{H}^{(0)} = \mathcal{H}_0 - \Re, \ \mathcal{H}_0 = \hat{H}_0 - \frac{1}{2k^2 n_0^2}\hat{W}_0,$$

$$\hat{H}_0 = -\frac{1}{2k^2 n_0^2}\nabla_\perp^2 + \frac{1}{2n_0^2}(n_0^2 - n^2),$$

$\hat{H}^{(1)}$ is the nonparaxial correction to the operator of paraxial propagation, which has the form

$$H^{(1)} = \frac{1}{2}\mathcal{H}_0^2 - \frac{1}{2}\Re^2, \ \Re = \frac{P}{2kn_0}.$$

The Hamiltonian $\hat{H}$ may be rewritten in terms of annihilation and creation operators in cylindrical coordinates [11]. This allows the matrix elements to be calculated analytically.

Consider a cylindrical waveguide with a parabolic distribution of the refractive index:

$$n^2(r) = n_0^2 - \omega^2 r^2, \qquad (5)$$

where $n_0$ is the refractive index on the waveguide axis, $\omega$ is the gradient parameter, $r = (x^2 + y^2)^{1/2}$.

In this case the solution of unperturbed equation $H_0 \Psi = \varepsilon_0 \Psi$ is described by radially symmetric Laguerre-Gauss functions:

$$\psi_{pl}(r,\varphi) = \left(\frac{k\omega}{\pi}\right)^{1/2}\left[\frac{p!}{(p+l)!}\right]^{1/2}(k\omega r^2)^{1/2}$$
$$\times \exp\left(-\frac{k\omega r^2}{2}\right)L_p^l(k\omega r^2)\exp(il\varphi) \qquad (6)$$

where $p$ and $l$ are the radial and azimuthal indices, accordingly, $l = v, v-2, v-4, \ldots 1$ or $0$.

The zero-order eigenvalues $\varepsilon_0$ are equal to $\varepsilon_0 = (\omega/kn_0^2)(2p+|l|+1) = (\omega/kn_0^2)(v+1)$, and the modes are $(v+1)$-fold degenerated, and for a given value of principal number $v$, the modes are degenerate on $l$. The degeneracy is due to the axial symmetry of the waveguide. If the polarization degeneracy of the modes is included then the modes are $2(v+1)$-fold degenerated.

The correction to the Hamiltonian describing spin-orbit interaction removes the polarization degeneracy. Indeed, the first order correction to the eigenvalue $\varepsilon^{(1)} = \langle vl\sigma | H^{(1)} | vl\sigma \rangle$ depends on the sign of the helicity and total angular momentum

$$\varepsilon^{(1)} = \frac{\eta^2}{32}\left[11(v+1)^2 - j^2 - 2j\cdot\sigma\right], \qquad (7)$$

where $\langle vl\sigma| = \langle\langle vl|, -i\sigma\langle vl|, e_z|$, $\eta = \omega/kn_0^2$ is the small parameter, $j = l + s$ is the total angular momentum, $\sigma$ is the helicity, and the longitudinal field component can be expressed through the transverse field components, i.e. $|e_z\rangle = (i/kn_0)\overline{\nabla}_\perp \bar{e}_\perp$.

Note that the first order correction does not remove the polarization degeneracy of modes completely. Similar result was obtained for the modes of a ring resonator made of a circular dielectric waveguide[6]. For the propagation constant correct to first order we have:

$$\beta_{vls} = kn_0\left\{1 - \eta(v+1) - \frac{\eta^2}{32}\left[11(v+1)^2 - j^2 - 2j\cdot\sigma\right]\right\}. \qquad (8)$$

The group delays of the modes are given by

$$\tau = \frac{z}{c}\frac{\partial\beta}{\partial k} \cong \frac{zn_0}{c} + \frac{zn_0}{c}\frac{\eta^2}{32}\left[11(v+1)^2 - j^2 - 2j\cdot\sigma\right], \qquad (9)$$

where $c$ is the light velocity, $z$ is the length of the fiber.

It is seen that the group velocities $v_g = z/\tau$ of vortex modes with right- and left-handed polarizations differ from each other, so the effective anisotropy is induced due to spin-orbit interaction. No such asymmetry exists in the case of zero orbital momentum $l = 0$.

Apart from polarization degeneracy, there is a degeneracy of modes with distinct orbital angular momentum and polarization but the same total angular momentum. The eigenvalues for modes can be analytically derived using a degenerate perturbation theory

$$\varepsilon_{1,2}^{(1)} = (\eta^2/32)\left[11(v+1)^2 - j^2 + 2j \right.$$
$$\left. \pm 10(v+1)\sqrt{(p+1)(p+l)}\right]. \qquad (10)$$

Degeneracy lifting $\delta\varepsilon = (5/8)\eta^2(v+1)\sqrt{(p+1)(p+l)}$ of the levels with the same $j$ occurs. Hybrid modes which are the solutions of (3) correspond to these levels

$$\psi_{vl}(r,\varphi) = a_{v,l,-1}\begin{pmatrix}|v,l\rangle \\ -i|v,l\rangle \\ e_{z1}\end{pmatrix} + b_{v,l,+1}\begin{pmatrix}|v,l-2\rangle \\ i|v,l-2\rangle \\ e_{z2}\end{pmatrix}. \qquad (11)$$

Hybrid polarization-OAM entangled states are the vector modes of a graded-index waveguide. Note that similar OAM mixing of LG modes was found in microcavities [5]. Because of the fact that the shift also takes place for the level with $l = 0$, its mechanism is not only connected with the spin-orbit interaction, but also with nonparaxiality and tensor interaction [12]. The degeneracy lifting might be viewed as an optical analogue of the Lamb shift, in which the level separation between the degenerate states with the same total angular momentum occurs.

In Fig.1a the relative propagation delay of modes compared with the fundamental mode as a function of radial mode number for various fundamental mode radiuses $w_0 = (2/k\omega)^{1/2}$ is presented. In Fig.1b the group delay splitting of the azimuthal modes of fixed radial indices ($p=0$) is presented. Note that the group delay splitting is one order of magnitude less than the relative propagation delay of modes.

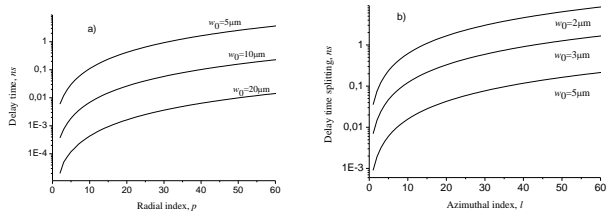

Fig.1. Delay time (a) and group delay splitting (b) as a function of radial and azimuthal indices, accordingly, $z = 1 km$, $n_0 = 1.5$.

LG beams are also the modes of the Fabry-Perot resonator and fiber ring resonators made of a circular graded-index waveguide. It is followed from the resonance condition that the frequency shift and the corresponding wavelength shift are equal to

$$\frac{\Delta f}{f} = \frac{\Delta\lambda}{\lambda} = \frac{11(v+1)^2 - j^2 - 2j\cdot\sigma}{8k^4 n_0^4 w_0^4}, \quad (12)$$

where $f = cn_0/\lambda$.

The same order of magnitude of the frequency shift was obtained in [9, 10]. The mode splitting of the azimuthal modes of fixed radial indices is given by $\Delta\lambda_l = \lambda_{l+1} - \lambda_l$. In Fig.2a the relative wavelength shifts as a function of azimuthal mode number for various beam radiuses are presented. The splitting increases with the azimuthal index $l$ and this behavior is highly sensitive to the size of the fundamental mode $w_0$. Similar result was received in nearly hemispherical microcavities experimentally [3]. In Fig.2b the relative degeneracy lifting $\delta\lambda_l/\lambda$ for various beam radiuses $w_0$ is presented. For the resonators with fundamental mode radius of the order of the light wavelength this split is about $1 nm$, which can be observed experimentally.

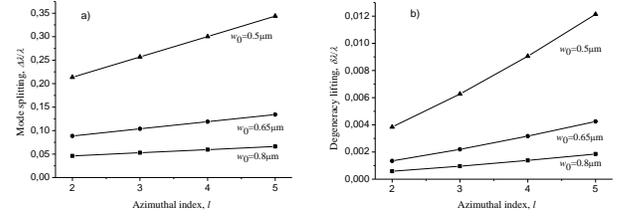

Fig.2. Mode splitting (a) and degeneracy lifting (b) as function of azimuthal index, $p=0, n_0 = 1.5$.

Thus, the frequency shift of modes in a graded-index fiber due to nonparaxiality and polarization effects is demonstrated. The degeneracy lifting of modes with distinct orbital angular momentum (OAM) and polarization but the same total angular momentum due to the spin-orbit (vector) and tensor forces is shown. This lifting is very small for conventional macroscopic optical fibers where $w_0 \gg \lambda$, but it becomes significant for fibers with diameter of the order of the light wavelength (Figs.1,2). The numerical estimations showed that the degeneracy lifting results in the delay time between degenerate modes of the order of $1 ns/km$ for an optical fiber with fundamental mode radius of the order of the light wavelength. Modes exhibiting hybrid entanglement between spin and orbital angular momentum may be useful for classical implementations of quantum communication and computational tasks [13].